\documentclass[3p,times]{elsarticle}

\usepackage{ecrc}

\volume{00}

\firstpage{1}

\journalname{Nuclear Physics A}

\runauth{}

\jid{nupha}

\usepackage{amssymb}
\usepackage{hyperref}

 \biboptions{sort&compress}

\usepackage{lineno}

\usepackage[figuresright]{rotating}

\newcommand{\pp}{pp}
\newcommand{\sqrts}{\sqrt{s}}
\newcommand{\sqrtsNN}{\sqrt{s_{\rm \scriptscriptstyle NN}}}
\newcommand{\PbPb}{\mbox{Pb--Pb}}

\newcommand{\GeV}{\mathrm{GeV}}

\newcommand{\gev}{\mathrm{GeV}}
\newcommand{\tev}{\mathrm{TeV}}

\newcommand{\RAA}{R_{\rm AA}}
\newcommand{\TAA}{T_{\rm AA}}

\newcommand{\pt}{p_{\rm T}}

\newcommand{\DtoKpi}{{\rm D^0\to K^-\pi^+}}
\newcommand{\DtoKpipi}{{\rm D^+\to K^-\pi^+\pi^+}}
\newcommand{\DstartoDpi}{{\rm D^{*+}\to D^0\pi^+}}
\newcommand{\Dzero}{{\rm D^0}}
\newcommand{\Dstar}{{\rm D^{*+}}}
\newcommand{\Dplus}{{\rm D^+}}

\begin{document}

\begin{frontmatter}



\dochead{}

\title{D mesons suppression in \PbPb~collisions at $\sqrtsNN = 2.76$~TeV measured by ALICE}


\author{Zaida Conesa del Valle for the ALICE Collaboration}

\address{European Organization for Nuclear Research (CERN), Geneva, Switzerland}

\begin{abstract}
%
The measurement of the prompt charm mesons $\Dzero$, $\Dplus$, $\Dstar$, and their antiparticles, in \PbPb~collisions at the LHC, at a centre-of-mass energy $\sqrtsNN=2.76~\tev$, with the ALICE detector, using 2010 data, is presented. 
The $\RAA$ of the three meson species show a suppression of a factor 3--4, for transverse momenta larger than $5~\gev/c$, in the 20\% most central collisions. The suppression is reduced for peripheral collisions. 
The results are compared with those of charged particles, charged pions, and non-prompt $J/\psi$, and also with theoretical calculations.
\end{abstract}


\end{frontmatter}


\section{Introduction}
\label{sec:Introduction}

Charm and beauty production in proton-proton collisions at the LHC is an important tool to test pQCD calculations in a new energy domain. 
Their spectra in heavy-ion interactions are influenced by the formation of hot and dense QCD matter. Due to their relative large mass and the difference between quark and gluon color charge, the medium effects for heavy flavor hadrons differ from those of light hadrons. RHIC results show that they effectively lose energy in the medium through (elastic and/or inelastic) interactions, but a possible quark-mass hierarchy has not yet been elucidated. LHC measurements at higher energies, with larger cross sections and better capabilities to separate charm and beauty production, can help to answer these questions. 

Open heavy flavor production can be measured with the ALICE experiment at the LHC in different colliding systems. 
The D meson reconstruction and analysis strategy using 2010 \PbPb~data at $\sqrtsNN=2.76$~TeV~\cite{ALICEDRaa,ALICEDpp7TeV,ALICEDpp276TeV} is summarized here. The results are compared with those of charged particles~\cite{ALICENchargedRaa}, charged pions, and non-prompt $J/\psi$~\cite{CMSquarkonia}, and also with theoretical calculations~\cite{eps09,vitev,vitevjet,whdg2011,horowitzAdSCFT,beraudo,gossiaux,bamps,cujet}. 

\section{Measurements in \PbPb~collisions at $\sqrtsNN=2.76$~TeV with 2010 data}
\label{sec:Analysis}

In heavy-ion collisions, the influence of the hot and dense QCD matter on the produced particles is commonly evaluated by comparing particle production in proton--proton (pp) and nucleus--nucleus (AA) collisions. It is usually presented in the form of the nuclear modification factor ($\RAA$):
\begin{equation}
\RAA(\pt) = \frac{1}{\langle \TAA \rangle} \cdot \frac{ {\rm d} N_{\rm AA}/{\rm d}\pt}{ {\rm d}\sigma_{\rm pp}/{\rm d}\pt } \, ,
\end{equation}
the ratio between the invariant yield in AA collisions (${\rm d} N_{\rm AA}/{\rm d}\pt$), and the pp cross section (${\rm d}\sigma_{\rm pp}/{\rm d}\pt$) normalized by the average nuclear overlap function\footnote{
The nuclear overlap function, $\TAA$, is defined as the convolution of the colliding ions nuclear density profiles, and it is evaluated in the Glauber model. 
The estimates of $\TAA$ in the Glauber model are in agreement with those computed in a Glauber Monte Carlo of the VZERO scintillators response, located at $2.8<\eta<5.1$ and $-3.7<\eta<-1.7$.
} ($\langle \TAA \rangle$). 


The data sample analyzed concerns minimum bias \PbPb~collisions at $\sqrtsNN=2.76$~TeV, collected in November and December 2010, with a $\mathcal{L}_{\rm int}=5$~nb$^{-1}$. 
$\Dzero$, $\Dplus$ and $\Dstar$ mesons, and their antiparticles, were reconstructed in the central rapidity region in their hadronic decay channels : $\DtoKpi$, $\DtoKpipi$, $\DstartoDpi$. 
Only good quality tracks, with $|\eta|<0.8$, at least 70 associated space points in the Time Projection Chamber (TPC), with $\chi^2/$ndf$ < 2$, and at least 2 (out of 6) associated hits in the Inner Tracking System (ITS), were selected to build the combinatorics. 
The selection of $\Dzero$ $(c\tau \sim123~\mu{\rm m})$ and $\Dplus$ $(c\tau \sim 312~\mu{\rm m})$ candidates was based on the reconstruction of their secondary vertex topologies. Typical selection variables were: the candidate decay length, the distance and angles of the decay products, and the minimum $\pt$ of the decay products. 
In the $\Dstar$ meson case, the decay $\Dzero$ secondary vertex topology was reconstructed. 
The selection cut values vary for each meson and $\pt$ bin, and were defined in order to have a large statistical significance of the signal and keep the selection efficiency as high as possible. 
The decay pions and kaons were selected using a $\pm 3\sigma$ cut around the expected energy deposit in the TPC and the time of flight in the TOF detectors\footnote{
In the $\Dstar$ case, for the 0-20\% centrality class, a $\pm 2\sigma$ cut was applied to the $\Dzero$ decay products.  
}. 
To build the $\Dstar$ candidates, no particle identification was required to the pions associated with $\Dzero$ candidates. 
%
%
Acceptance and reconstruction efficiencies were evaluated with Monte Carlo simulations of minimum-bias \PbPb~collisions produced with the HIJING v1.36 event generator. Prompt and feed-down (B decays) D meson signals were added using \pp~events from the PYTHIA v6.4.21 event generator with the Perugia-0 tuning~\cite{ALICEDRaa}.
A detailed description of the apparatus geometry and response, the experimental conditions, and their evolution with time was included. 
%
Prompt D meson yields were obtained by subtracting the contribution of D mesons from B decays. It was evaluated using the beauty cross sections from FONLL~\cite{fonll} calculations, the B$\longrightarrow$D decay kinematics from the EvtGen package~\cite{evtgen} and the Monte Carlo efficiencies for feed-down D mesons. This contribution was renormalized by the average nuclear overlap function in each centrality class, and the nuclear modification factor of feed-down D mesons ($\RAA^{\rm feed-down}$). 
To perform the correction, it was assumed that $\RAA^{\rm feed-down}=\RAA^{\rm prompt}$, while $\RAA^{\rm feed-down}$ was allowed to vary to evaluate the measurement systematics. 
%
More details on the analysis procedure and the systematic uncertainties determination can be found in ref.~\cite{ALICEDRaa}. 


$\Dzero$, $\Dplus$ and $\Dstar$ mesons cross sections in \pp~collisions at $\sqrts=7$~TeV and $2.76$~TeV were reported in ref.~\cite{ALICEDpp7TeV,ALICEDpp276TeV}. Due to the limited statistics of the  $\sqrts=2.76$~TeV data sample, the $7$~TeV cross sections scaled down to $2.76$~TeV were used as proton-proton reference. The scaling factor was evaluated as the ratio of the FONLL~\cite{fonll} cross sections at the two energies. The scaling factor uncertainties include the FONLL calculation uncertainties: $0.5 m_{\rm T}<\mu_{R}<2 m_{\rm T}$, $0.5 m_{\rm T}<\mu_F<2 m_{\rm T}$, $1.3 < m_{\rm c} < 1.7~\GeV/c^2$, where $\mu_F=\mu_R=m_{\rm T}$ and $m_{\rm T}= \sqrt{\pt^2 +m^2_{\rm c}}$. 
The scaling procedure was validated by comparing the scaled results to the measurements at $2.76$~TeV~\cite{ALICEDpp276TeV}, and by scaling the $7$~TeV data to Tevatron energies and comparing to CDF data~\cite{scaling}. The scaling was also verified by using GM-VFNS calculations~\cite{gmvfns}.


\section{Results}
\label{sec:Results}

\begin{figure}[!htbp]  
\begin{center}        
\includegraphics[width=0.85\textwidth]{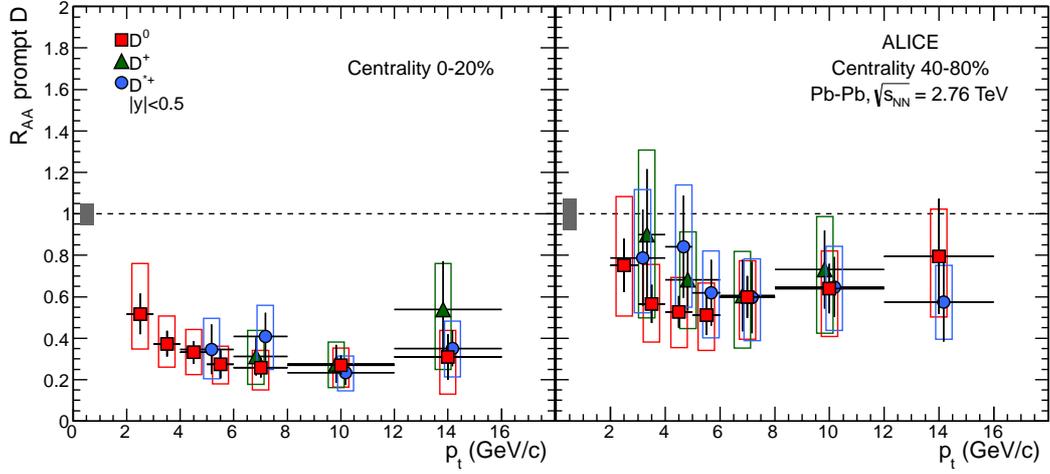}
\caption{
$\RAA$ for prompt $\Dzero$, $\Dplus$ and $\Dstar$ in the 0--20\% (left) and 40--80\% (right) centrality classes~\cite{ALICEDRaa}. Statistical (bars), systematic (empty boxes), and normalization (full box) uncertainties are shown. Horizontal error bars reflect bin widths, symbols were placed at the centre of the bin.
}
\label{fig:RaaDmesons}
\end{center}
\end{figure}

\begin{figure}[!htbp]  
\begin{center}        
\includegraphics[width=0.475\textwidth]{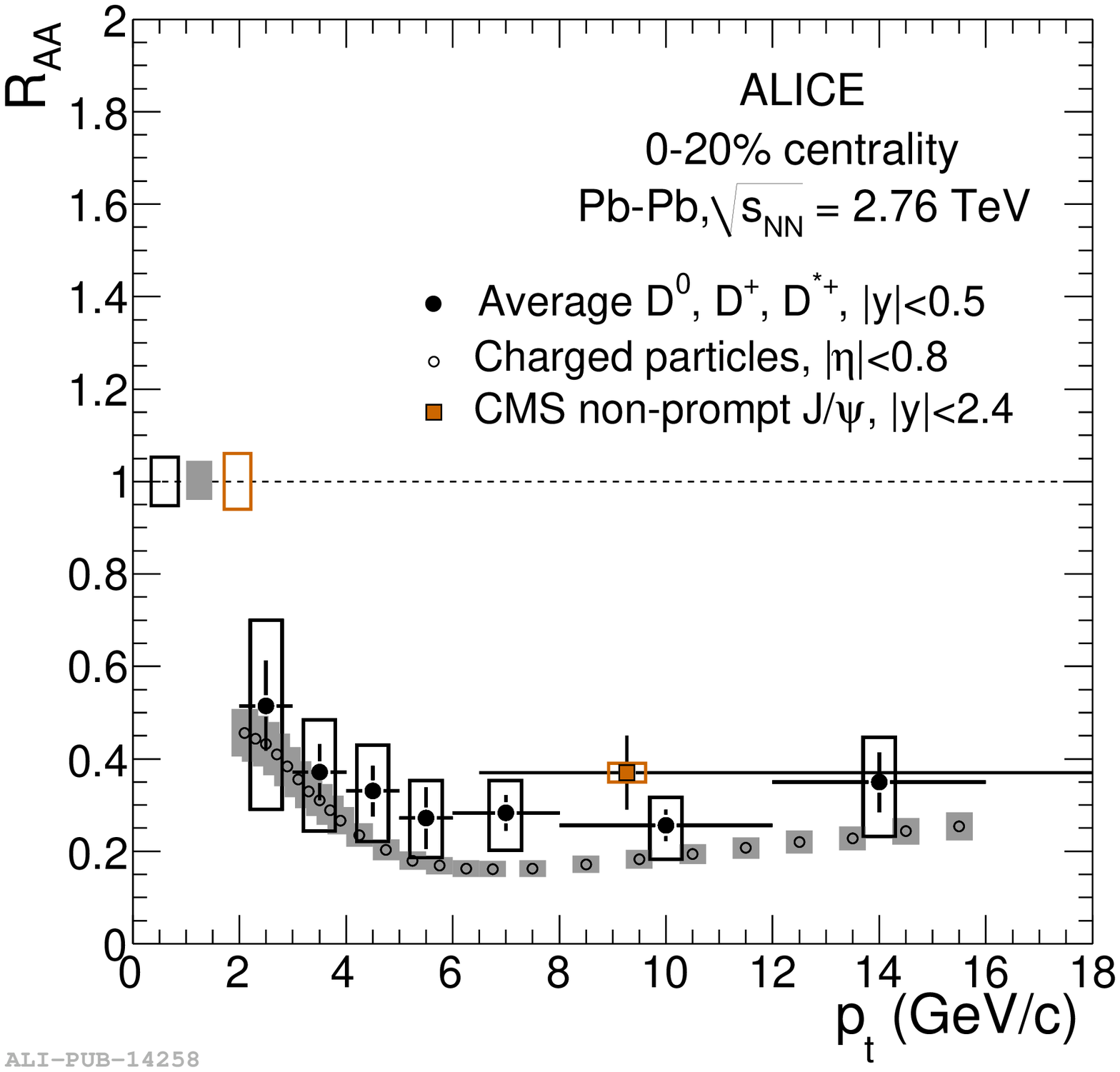}
\includegraphics[width=0.46\textwidth]{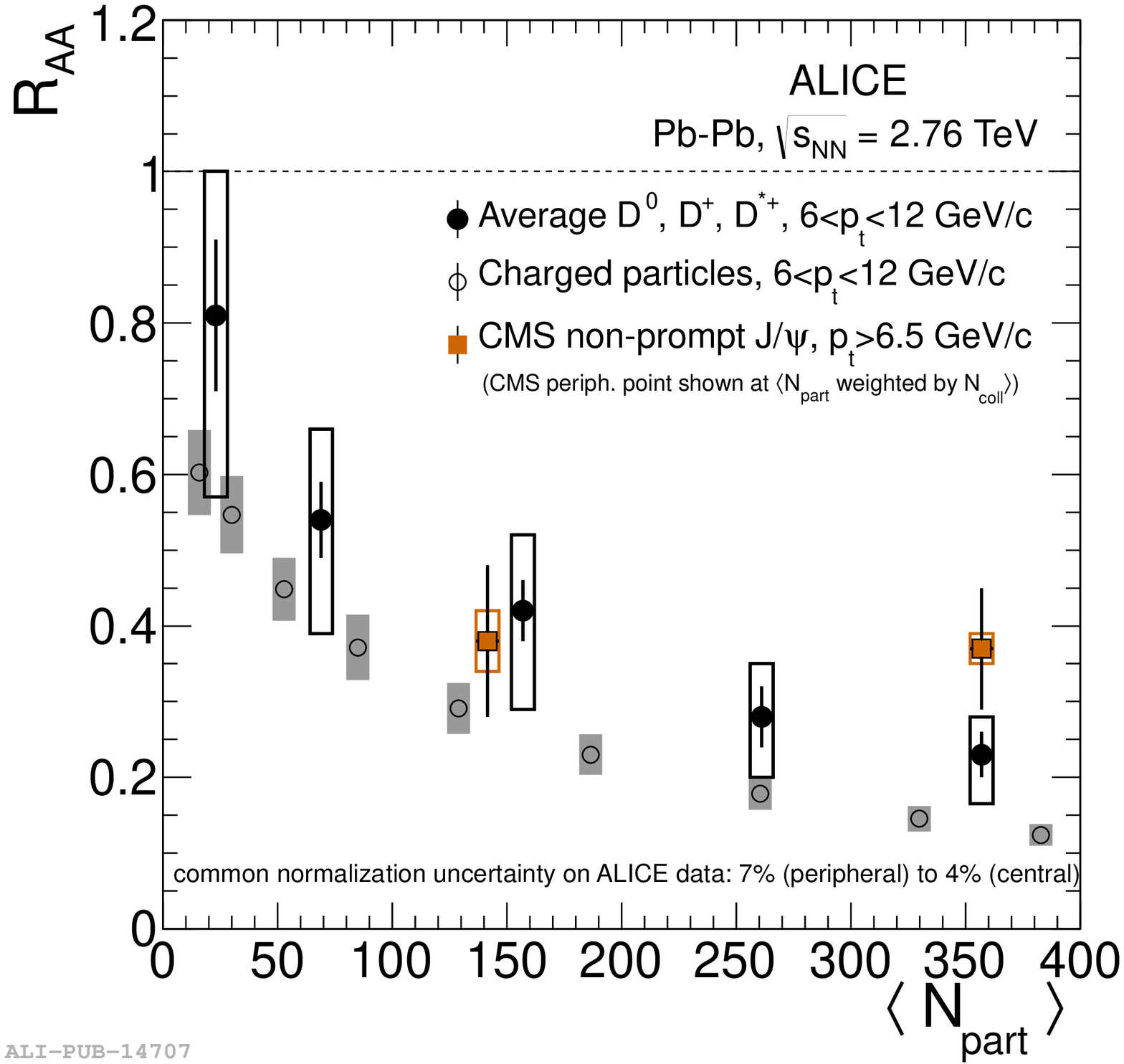}
\caption{
Left: Average D meson ($\Dzero$, $\Dplus$ and $\Dstar$) $\RAA$ in the 0--20\% centrality class vs. $\pt$ compared with charged particles and non-prompt $J/\psi$~\cite{ALICEDRaa}.
Right: Average D meson $\RAA$ vs. centrality, represented by the average number of participant nucleons, $\langle N_{\rm part} \rangle$, compared with charged particles and CMS non-prompt $J/\psi$.
}
\label{fig:AverageDvsPt}
\label{fig:AverageDvsNpart}
\end{center}
\end{figure}

The transverse momentum distributions ${\rm d}N/{\rm d}\pt$ and $\RAA(\pt)$ of prompt $\Dzero$, $\Dplus$ and $\Dstar$ mesons in the 0--20\% and 40--80\% centrality classes were first presented in ref.~\cite{ALICEDRaa}. Figure~\ref{fig:RaaDmesons} shows $\RAA(\pt)$ of prompt D mesons. 
The results of the three D meson species are in agreement within statistical uncertainties and show a suppression by a factor of 3--4 for $\pt>5~\GeV/c$ in the 0--20\% centrality class. 
The centrality dependence was studied in more detail in wider $\pt$ bins and thinner centrality bins and reported in ref.~\cite{ALICEDRaa}. A tendency to have a larger suppression in the most central collisions is observed, see also Fig.~\ref{fig:AverageDvsNpart}~(right). 

The average D meson $\RAA(\pt)$ was computed from the weighted average\footnote{
Statistical uncertainties were used as weights to evaluate the  average D meson $\RAA(\pt)$.
} of $\Dzero$, $\Dplus$ and $\Dstar$ $\RAA(\pt)$. 
%
The average D meson $\RAA$ is compared in Fig.~\ref{fig:AverageDvsPt} with charged particles~\cite{ALICENchargedRaa} and non-prompt $J/\psi$ from CMS~\cite{CMSquarkonia}. 
$\RAA(\pt)$ of the average D meson and charged particles~\cite{ALICENchargedRaa} are similar in magnitude and $\pt$ trend, while non-prompt $J/\psi$ shows a smaller suppression than that of charged particles. 
This observation can also be done in Fig.~\ref{fig:AverageDvsNpart}~(right), which displays $\RAA$ vs. centrality, represented by the average number of participant nucleons $\langle N_{\rm part} \rangle$, for $\pt>6~\gev/c$. 
However the comparison of non-prompt $J/\psi$ and the average D meson $\RAA$ is not yet conclusive and requires more precise and differential measurements.

\begin{figure}[!htbp]  
\begin{center}        
\includegraphics[width=0.47\textwidth]{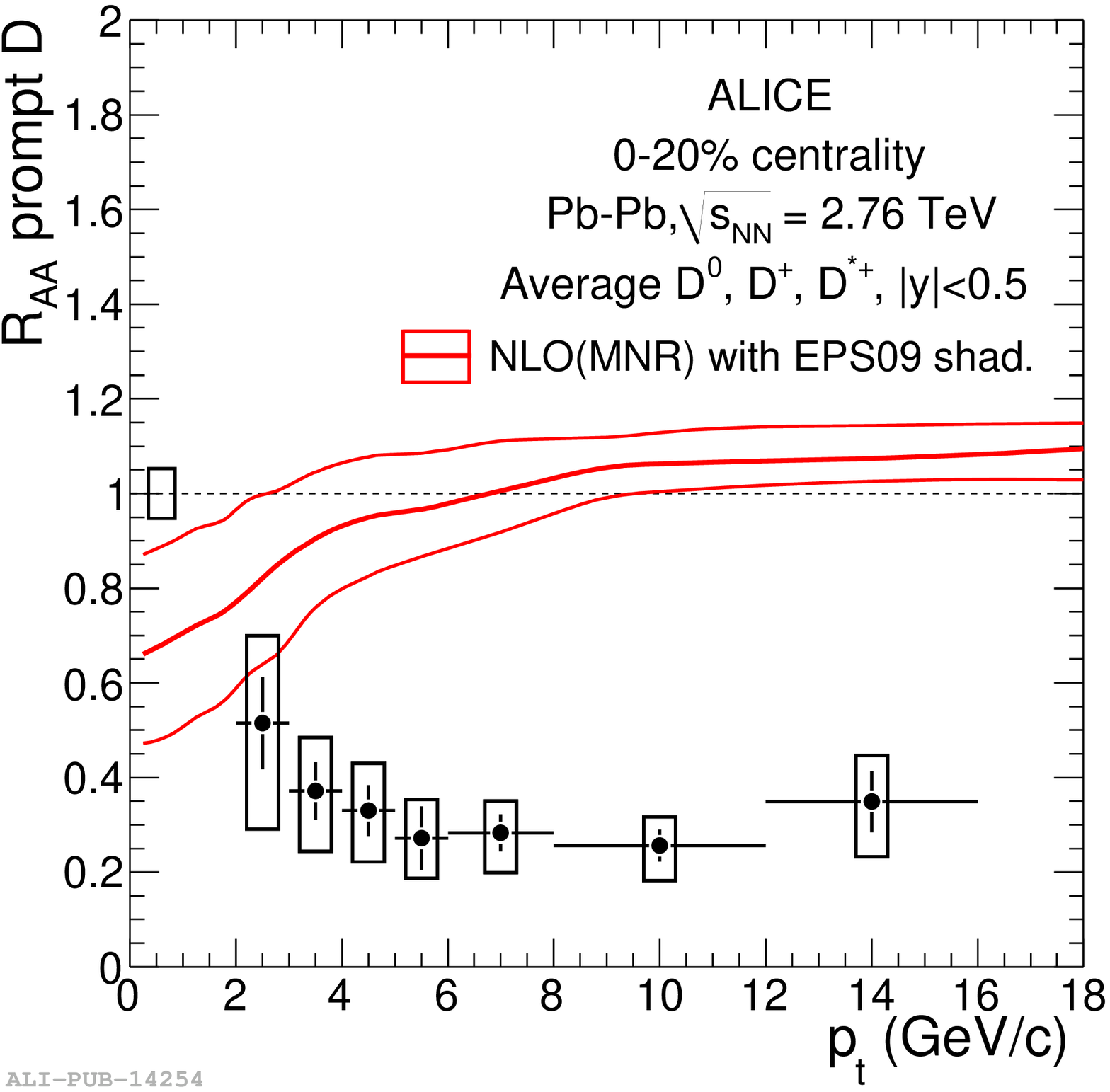}
\includegraphics[width=0.48\textwidth]{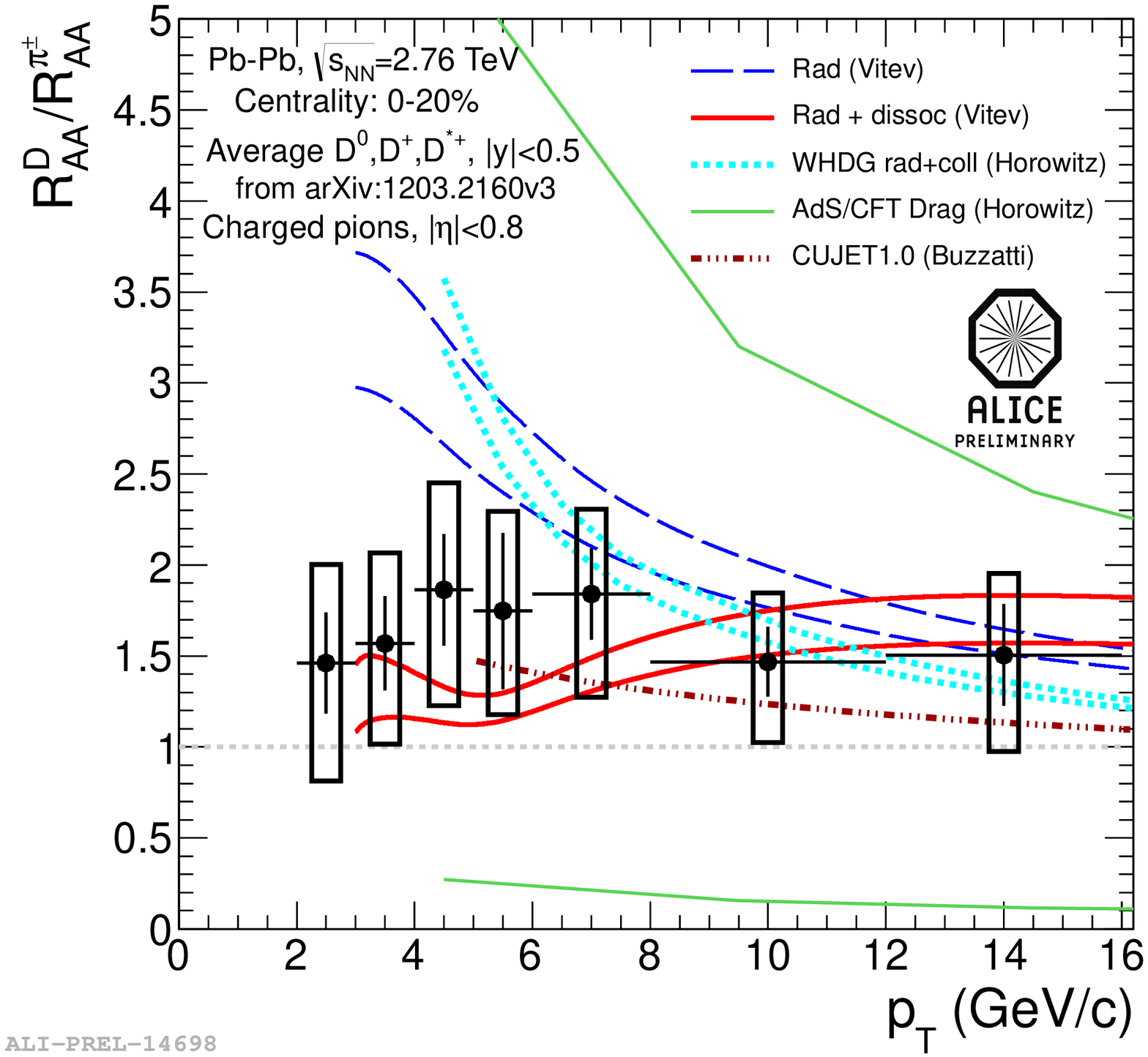}
\caption{
Left: Average D meson ($\Dzero$, $\Dplus$ and $\Dstar$) $\RAA$ in the 0--20\% centrality class vs. $\pt$ compared with NLO (MNR) calculations~\cite{MNR} with EPS09~\cite{eps09} shadowing~\cite{ALICEDRaa}.
Right: Ratio of the average D meson and charged pion $\RAA$ vs $\pt$ compared with theoretical calculations~\cite{vitev,vitevjet,whdg2011,horowitzAdSCFT,cujet}.
}
\label{fig:AverageDvsPtShad}
\label{fig:RaaDvsRaaPion}
\end{center}
\end{figure}

The results in the 0--20\% centrality class are compared in Fig.~\ref{fig:AverageDvsPtShad}~(left) to the NLO (MNR) calculations~\cite{MNR} with EPS09~\cite{eps09} shadowing. The magnitude of D meson suppression at $\pt>5~\GeV/c$ in the most central class can therefore not be explained by initial-state effects. 
Figure~\ref{fig:RaaDvsRaaPion}~(right) presents the ratio of the average D mesons and charged pions $\RAA(\pt)$ in 0--20\% centrality class. This ratio shows no clear $\pt$ dependence and is of about 1.5. 
Several theoretical models based on parton energy loss compute the charm nuclear modification factor~\cite{vitev,vitevjet,whdg2011,horowitzAdSCFT,beraudo,gossiaux,bamps,cujet}. 
Those having results for both D mesons and charged pions are shown in Fig.~\ref{fig:RaaDvsRaaPion}~(right)~\cite{vitev,vitevjet,whdg2011,horowitzAdSCFT,cujet}. 
All of them describe reasonably well this ratio and the D mesons and charged pions $\RAA(\pt)$. However, it can be noted that a model based on the AdS/CFT correspondence seems to underestimate charm $\RAA$ and has limited predictive power for light hadron $\RAA$. 



\vspace{-12pt}

\bibliographystyle{elsarticle-num}



\end{document}